
%
%
%

\input harvmac
\def\t{\theta}
\def\L{\Lambda}
\def\ajou#1&#2(#3){\ \sl#1\bf#2\rm(19#3)}

\Title{CTP\# 2181}
{Relating Black Holes in Two and Three Dimensions$^*$\footnote{}{$^*$
This work is
supported in part by funds provided by the U. S. Department of energy
(D.O.E) under contract \#DE-AC02-76ER03069, and by the NSF under grant
PHY-9111188.}}

\centerline{Ana Ach\'ucarro\footnote{$^\dagger$}
{aachucar@pearl.tufts.edu}}
\centerline{\sl Dept. of Mathematics, Tufts University,
Medford, MA 02174, USA.}
\centerline{\sl and}
\centerline{\sl Dept. of Theoretical Physics,
University of the Basque Country, Bilbao, Spain.}
\smallskip
\centerline{and}
\smallskip
\centerline{Miguel E. Ortiz\footnote{$^\ddagger$}{ortiz@mitlns.mit.edu}}
\centerline{\sl Center for Theoretical Physics,}
\centerline{\sl Laboratory for Nuclear Science and Department of Physics,}
\centerline{\sl Massachusetts Institute of Technology,
Cambridge, MA 02139, USA.}
\vskip 0.2in
The three dimensional black hole solutions of
Ba\~nados, Teitelboim and Zanelli (BTZ) are dimensionally reduced
in various different ways.
Solutions are obtained to the Jackiw-Teitelboim
theory of two dimensional gravity for spinless BTZ black holes,
and to a simple
extension with a non-zero dilaton potential for black holes of fixed spin.
Similar reductions are given for charged black holes. The resulting
two dimensional solutions are themselves black holes, and
are appropriate for investigating exact ``S-wave'' scattering in
the BTZ metrics. Using a different dimensional reduction to the string
inspired model of two dimensional gravity, the BTZ solutions are related
to the familiar two dimensional black hole and the linear dilaton vacuum.
\Date{January 1993}

\eject

Although examples of black hole solutions abound in two and four
dimensions, it was until recently believed that no such solutions exist
in three spacetime dimensions\ref\djt{Solutions of pure Einstein gravity
have been classified in S. Deser, R. Jackiw and G. 't Hooft, \ajou
Ann. Phys. & 152 (84) 220, and no black hole solutions were found. It is
straightforward, if tedious, to see that no analogues exist in three
dimensions of the dilatonic black hole solutions of G. Gibbons and
K. Maeda, \ajou Nucl.
Phys. & B298 (88) 741, and
D. Garfinkle, G. Horowitz, and A.
Strominger, \ajou Phys. Rev. &D43 (91) 3140; Erratum: \ajou Phys. Rev.
&D45 (92) 3888. Charged solutions with a horizon
in 2+1 dimensions are known, but they do not conform to the usual
notion of a black hole;
see the discussion in
B. Reznik, Phys. Rev. {\bf D45} 2151 (1992).}.
However, in a recent paper, Ba\~nados,
Teitelboim and Zanelli (BTZ) \ref\btz{M. Ba\~nados, C. Teitelboim and J.
Zanelli, \ajou Phys. Rev. Lett. & 69 (92) 1849},
\ref\bhtz{M. Ba\~nados, M. Henneaux,
C. Teitelboim and J. Zanelli, ``Geometry of the 2+1 black hole'', Santiago
preprint (1993).}
found a vacuum solution to Einstein gravity with a negative
cosmological constant which may be interpreted as a black hole. The
solution has everywhere constant curvature, but the global topology is
different to that of three dimensional Anti de Sitter space. As a result, the
causal structure of the solution is closer to that of the Schwarzschild
solution. However, the singularity hidden behind the horizon is of a
weaker form than that of Schwarzschild \bhtz,
\ref\clm{D. Cangemi, M. Leblanc and R. Mann, ``Gauge
formulation of the spinning black hole in 2+1 dimensional Anti-De Sitter
space'', MIT preprint CTP \#2162, gr-qc/9211013 (1992).}.

Below we discuss how the BTZ solutions may be dimensionally reduced to
solutions of various two dimensional theories of gravity. Our motivation is
provided by the recent evidence that progress may be made in
understanding black hole radiation and evaporation in the context of two
dimensions \ref\hs{See for example J. A. Harvey and A. Strominger, ``Quantum
aspects of black holes'', U. of Chicago preprint EFI-92-41,
hep-th/9209055 (1992), and references therein.}.
The solutions we derive may all be interpreted as two
dimensional black holes, and some of
the corresponding two dimensional theories of
gravity may in principle be used to exactly describe the
scattering of rotationally
symmetric matter (``S-waves'')
off the three dimensional black holes. Since our present understanding of
three dimensional quantum gravity coupled to matter indicates that it is
non-renormaliseable,
these models
may provide the only route for understanding the quantum behaviour
of the BTZ solutions.

\bigskip

First let us review the
BTZ solutions \btz, \bhtz, \clm. They arise in a three dimensional
theory of gravity
\eqn\two{S=\int d^3x \sqrt{-g}\left(R+2\L\right)}
with a negative cosmological constant, {\it i.e.} $\L>0$.
It is straightforward to
check that the Einstein field equations
\eqn\five{R_{\mu\nu}-{1\over 2}Rg_{\mu\nu}+\L g_{\mu\nu}=0}
are solved by the metric
\eqn\six{ds^2=-N^2(r)dt^2+{dr^2\over N^2(r)}+r^2\left(N^\t(r)dt
+d\t\right)^2}
where
\eqn\seven{N^2(r)=\L r^2-M+{J^2\over 4r^2},\qquad N^\t(r)=-{J\over 2r^2}.}

For $J=0$, this metric is similar to Schwarzschild.
It has a single horizon at $r=\sqrt{M/\L}$,
and a singularity at $r=0$. However there are two important differences.
Firstly, \six~is not asymptotically flat -- it is a constant curvature
metric. Secondly, the singularity at $r=0$ is much weaker than that of
Schwarzschild spacetime. Whereas the singularity of Schwarzschild is
manifested by the power law divergence of curvature scalars at small $r$,
the BTZ solution has at most a delta function singularity at $r=0$, since
everywhere else the spacetime is of constant curvature \ref\sing{The
singularity at $r=0$ arises because \six~is the covering space of AdS$^3$,
identified by a discrete subgroup of the isometry group. It is due to
a singularity in the causal structure, but it is unclear whether it should
be regarded as a conical singularity in the usual sense \bhtz, \clm.}.
An interesting special case occurs when $M=0$. In this
case the metric takes the form
\eqn\eight{ds^2=-\L r^2dt^2+{dr^2\over \L r^2}+r^2d\t^2.}
This should be regarded as the extremal or vacuum solution of the $J=0$ family.
The spatial geometry of this vacuum solution is an infinite wormhole, whose
radius shrinks to zero at $r=0$
at an infinite spacelike proper distance from the
asymptotic region. Geodesics reach the end of the wormhole at
$r=0$ within a finite proper time, and it is unclear how they should be
continued beyond $r=0$. However, since
this solution is extremal, the expectation is that a horizon develops
before even the lightest test particle reaches this point.

If $J\ne 0$, \six~has two horizons, and its causal structure is similar to
the Reissner-Nordstr\"om spacetime. When $M=\sqrt{\L}|J|$, the two horizons
coincide, and this should be regarded as the vacuum solution for fixed
$J$. As before, the spacetimes are of constant curvature and have a
singularity at $r=0$, with at most distributional torsion and
curvature at that point \bhtz, \clm.
Solutions with $M<\sqrt{\L}|J|$ have no
horizon. It has been conjectured in \btz~that
these should be discarded since they
contain a naked singularity, except when $M=-1$, $J=0$, when the spacetime
is just AdS$_3$.

The thermodynamics of the BTZ black holes suggest that evaporation of the
black holes should take place. Since the temperature decreases with
decreasing mass, and is zero for the extremal solutions, these seem to be
the natural endpoints of evaporation \btz.

\bigskip

Let us now discuss the possible dimensional reductions of the BTZ black
holes.
We begin with the straightforward dimensional reduction from Einstein
gravity with a cosmological constant in three dimensions to the
Jackiw-Teitelboim (JT) theory
\ref\jt{R. Jackiw, in {\sl Quantum Theory of Gravity}, ed. S. Christensen,
Hilger, Bristol, 1984; C. Teitelboim, {\sl ibid}.} in two.
Suppose that the gravitational
field in three dimensions is independent of a single co-ordinate, which we
shall call $\t$, and that the metric may be written in the form,
\eqn\one{ds^2=g_{\mu\nu}dx^\mu
		dx^\nu=h_{ij}(x^i)dx^idx^j+\Phi^2(x^i)d\theta^2}
where $\mu,\nu=0,1,2$ and $i,j=0,1$.
Then it is simple to see that the action \two~reduces to
\eqn\three{S=\int d^2x \sqrt{-h}\Phi\left(R+2\L\right)}
which is precisely the JT action. It follows that
any solution to the equations following from \two, of the form \one, yields
a solution
\eqn\four{h_{ij}(x^i),\qquad \Phi(x^i)}
of the equations following from \three. We shall henceforth refer to the
field $\Phi$ as the dilaton.

The BTZ solutions with $J=0$ are of the form \one, and they
therefore yield a solution to the JT model~\ref\jc{These
may be related to
the general solutions given in R. Jackiw, ``Gauge theories for
gravity on a line'', MIT preprint CTP\# 2105, hep-th/9206093 (1992).
The parameter $M$
corresponds to the parameter $\alpha_0$ of his equation (20),
and the other two parameters
$\alpha_a$ correspond to a shift of origin not explicit in our solution.
Although these parameters can be absorbed into the metric by
a change in co-ordinates, they cannot be made to simultaneously disappear
from both the metric and the dilaton.},
\eqn\ten{ds^2=-(\L r^2-M)dt^2+{dr^2\over \L r^2-M}\; ,\qquad \Phi=r.}
It is important to note that the dimensional reduction, although trivial in
appearance, has radically changed the properties
of the metric. In three
dimensions, the point $r=0$ is singular. The two dimensional metric is
perfectly well-behaved at $r=0$.
This is a result of the weak nature of the singularity in the
three dimensional solution.
The metric \ten~may be analytically extended beyond $r=0$,
using the co-ordinate transformation
\eqn\fifty{r=\sqrt{M\over \L}\cosh\rho\sin\left(\sqrt{\L}\tau\right),\qquad
\tanh\left(\sqrt{\L
M}t\right)={\tanh\rho\over\cos\left(\sqrt{\L}\tau\right)}}
and the
maximally extended spacetime is the whole of two dimensional
Anti-De Sitter spacetime,
\eqn\sixty{ds^2=-\cosh^2\rho d\tau^2+\L^{-1}d\rho^2,\qquad
-\infty<\tau,\rho< \infty,}
which of course has no horizons. The dilaton in these co-ordinates takes
the form
\eqn\abc{\Phi=\sqrt{M\over \L}\cosh\rho\sin\left(\sqrt{\L}\tau\right),}
which vanishes when $\sqrt{\L}\tau=n\pi$. The
embeddings of \ten~into \sixty~ are shown in Fig. 1.

In order
to interpret the two dimensional solution as a black hole, we must look at
the behaviour of the dilaton. Recall that the dilaton is the $\t \t$
component of the three dimensional metric, and that  the three dimensional
solution is singular where it vanishes. It is therefore natural to cut the
two dimensional spacetime off at this point, which we call the
``strong coupling'' region (although this name should not be taken too
literally), if we wish to use the JT theory to model
three dimensional physics. In this case, \ten~does indeed represent a black
hole, whose Penrose diagram, shown in Fig. 2, is identical to that of its
three dimensional counterpart.

The two dimensional version of the extremal
solution is also not geodesically complete unless it is extended (see also
Ref. \ref\mike{M. Crescimanno, ``Equilibrium two dimensional dilatonic
spacetimes'', MIT preprint CTP \#2174 (1993).} for a discussion of this
spacetime). The extended version of the spacetime is also identical to
Anti-De Sitter spacetime,
but for our purposes,
we should again restrict our attention to the
region where the dilaton is greater than zero.
In this extremal case the spacetime
\eqn\thirty{ds^2=-\L r^2dt^2+{dr^2\over \L r^2}}
has a ``naked singularity'' at $r=0$ in the sense that the region where
$\Phi=0$ is not hidden behind a horizon. However, this is of no
consequence: In both two and three dimensions (as discussed above)
we expect that no matter
can probe this region of spacetime without a horizon developing.
A Penrose diagram of the restricted extremal solution is shown in Fig. 3,
which again is identical to the three dimensional Penrose diagram.
Incidentally, this region is
the Steady State Universe solution \ref\he{S. W. Hawking and G. F. R.
Ellis, {\it The large scale structure of space-time}, CUP, Cambridge
(1973).}, but with timelike and spacelike directions interchanged.

The two dimensional reduction outlined above yields a two dimensional
theory which can be used to model S-wave scattering off a spinless
BTZ black hole. That solutions in
two dimensions must be
restricted by hand to end where the dilaton vanishes is both a
blessing
and a curse. On the one hand, there is no singular region to worry
about, but on the other we must specify boundary conditions at $r=0$
in some fashion.
In principle, however, by coupling matter to \three~in the
natural way for a dimensionally reduced theory,
\eqn\fourteen{S=\int d^2x\sqrt{-h}\Phi\left(R+2\L+h^{ij}
\partial_if\partial_jf\right),}
and imposing an appropriate boundary condition at $\Phi=0$,
it may be possible to model the evaporation of a spinless
BTZ black hole.

It is also possible to construct an effective two dimensional theory which
arises from the dimensional reduction of the $J\ne 0$ solutions of BTZ, and
which can be used to study S-wave scattering for the spinning black holes.
We begin by considering the reduction of a metric of the form
\eqn\eleven{ds^2=h_{ij}(x^i)dx^idx^j+\Phi^2(x^i)\left(d\t
+A_i(x^i)dx^i\right)^2.}
The corresponding two dimensional theory involves the three fields
$h_{ij}$, $A_i$ and $\Phi$. If we wish to consider spacetimes of fixed
spin, however, we can use the following identity,
\eqn\twelve{{\Phi^3\epsilon^{ij}\partial_iA_j\over\sqrt{-h}}={\rm
constant}}
which follows from the field equation for $A_i$.
The constant is precisely the spin $J$ of the metric \eleven, since it is
equal to the charge corresponding to asymptotic rotational invariance
\ref\brown{The charge is given in J. D. Brown and M. Henneaux,
\ajou Comm. Math. Phys. & 104 (86) 207, as $2\pi
J[g_{\alpha\beta},\pi^{\alpha\beta}]=\lim_{r\to\infty}\oint dS_\alpha\,
2\xi^\beta{\pi^\alpha}_\beta$, where $\xi=\partial/\partial\t$,
$\pi^{\alpha\beta}$ is
the usual momentum conjugate to $g_{\alpha\beta}$ in canonical gravity, and
the integral is taken over the boundary of a spacelike hypersurface. For a
metric of the form \eleven, the integrand over a circle at infinity is
precisely the constant expression in \twelve. Note also that \twelve~says
that the integrand of the surface term is constant over the entire
spacelike hypersurface of constant $x^0$; by Stokes's theorem it appears at
first sight that $J$ should therefore be zero. However, the contribution
giving rise to $J$ comes from a distributional source for angular momentum
at the singularity of the three dimensional spacetime as could be
expected.}.
Using this identity, the action \one~for spacetimes of spin $J$
may be dimensionally reduced to
\eqn\thirteen{S=\int d^2x\sqrt{-h}\Phi\left(R+2\L-{J\over 2\Phi^4}\right).}
This is an appropriate effective action for looking at S-wave scattering off
a spinning BTZ black hole (interaction with rotationally symmetric matter
will of course keep the black hole in the spin $J$ sector).
Again, any solution $h_{ij}$, $\Phi$ to \thirteen~corresponds to a solution
to \two~of spin $J$, of the form \eleven.
In particular, the $t,r$ section of the BTZ black hole of spin $J$
\eqn\forty{ds^2=-\left(\L r^2-M+{J^2\over 4r^2}\right)dt^2+{dr^2\over
\left(\L r^2-M+{J^2\over 4r^2}\right)}}
is a solution to
\thirteen, with $\Phi=r$. However, in this case, the equation for the
scalar curvature is
\eqn\twenty{R+2\L+{3J\over 2\Phi^4}=0}
so that $R$ need not be constant. Indeed the two dimensional spinning
black hole \forty~and extremal solution have power law
singularities in $R$ at $r=0$.
The Penrose diagram for each of these spacetimes is identical to that of
the three dimensional metric, and is shown in Fig. 4.

In addition to the BTZ solutions described above, it was also shown in
\btz~that charged black hole solutions similar to \six~exist.
These are solutions following from the action
\eqn\acbh{
\int d^3x \sqrt{-g}\left(R+2\L+{4\pi G}F_{\mu\nu}F^{\mu\nu}
\right)}
and take the form \six, but with
\eqn\xc{
N^2(r)=\L r^2-M-8\pi G Q^2\ln(r/r_0)+{J^2\over 4r^2}\qquad{\rm
and}\qquad F_{rt}=-{Q\over r}.
}
These solutions have a power law curvature singularity at $r=0$, where
$R\sim 8\pi G Q^2/r^2$. They
can have two, one or no horizons, depending on the relative
values of $\L$, $J$, $GQ^2$, and $M'=M-8\pi GQ^2\ln (\sqrt{\L}r_0)$.
In the simplest case $J=0$, these possibilities depend on whether
\eqn\xb{
M'-4\pi GQ^2\left(1-\ln\left[{4\pi GQ^2}\right]\right)
}
is greater than, equal to, or less than zero respectively.

The action \acbh~may be dimensionally reduced in a similar way to
that described above, in both the $J=0$ and the $J\ne 0$ sectors, provided
that we assume that $F_{rt}$ is independent of $\t$ and that the other
two components $F_{\mu\t}$ vanish. The
resulting two dimensional action is
\eqn\xa{
\int d^2x \sqrt{-h} \Phi\left(R+2\L-{J\over 2\Phi^4}+4\pi G
F_{\mu\nu}F^{\mu\nu}\right)}
The solution in two dimensions corresponding to the BTZ solution is the
obvious analogue
of \ten~and \forty~for $Q\ne 0$. The electromagnetic field has
$F_{rt}={-Q/r}$ as before. This two dimensional spacetime
has a curvature singularity at $r=0$, even for $J=0$,
since $R=-2\L -8\pi GQ^2/r^2-3J^2/2r^4$. As in three dimensions, this may
be a naked singularity or may be shielded by one or two horizons.

\bigskip

Finally, let us describe a third dimensional reduction of the
uncharged
BTZ black holes. This involves the reduction introduced in \ref\ana{A.
Ach\'ucarro, ``Lineal gravity from planar gravity'', Tufts preprint,
hep-th/9207108 (1992), to appear in Phys. Rev. Lett.}
from the three dimensional action \two~to the
string inspired action for two dimensional gravity \hs, \ref\wbh{E. Witten,
\ajou Phys. Rev. & D44 (91) 314; G. Mandal, A. Sengupta and S. Wadia, \ajou
Mod. Phys. Lett. & A6 (91) 1685.},
\eqn\zzz{\int d^2x \sqrt{-\hat{h}} \, e^{-2\phi}\left(\hat{R}+4\left(
\nabla\phi\right)^2+2\lambda\right)}
or
rather, to the action \ref\hv{H. Verlinde, proceedings of the Sixth Marcel
Grossmann Meeting, to be published, (1992).}
\eqn\aa{\int d^2 x \sqrt{-h} \, (\Phi R + 2\lambda ),}
which is obtained from the string-inspired action by means of the
identification\break ${\hat h}_{ij} = h_{ij}e^{2 \phi}$, $\Phi =
e^{-2\phi} $. To get the action \aa~from \two, consider the usual
Kaluza-Klein dimensional reduction used above to obtain the JT action
\three~(or
the equivalent action obtained from spinning metrics \eleven)
followed by a shift of $\Phi$ by a constant:
\eqn\bb{\Phi \rightarrow \Phi + {\lambda \over \Lambda},}
(this procedure
is in fact equivalent to implementing the shift \bb~on the action
\two~and then reducing \ana).
It yields a two-parameter family of actions  which include
the string-inspired action in the limit $\Lambda \rightarrow 0$.
Notice that the potentially divergent term
\eqn\cc{\int d^2x {\lambda \over \Lambda}R}
is proportional
to the Euler characteristic of the two dimensional manifold and, being a
topological invariant, it does not affect the equations of motion. Also, note
that the effect of the shift as seen from the three-dimensional point of view
is to push all points in
the ``extra'' $\t$ dimension an infinite distance away,
making the proper length of a $\t$ orbit diverge.

Unlike the JT reductions, this procedure
gives the same result irrespective of
the value of $J$. This  is best understood in the first order
formalism, where the dreibein $e^a$ and the spin connection $\omega^a$
are the dynamical variables (the action \two~is then replaced by
the Chern-Simons action for
the Anti de Sitter
group SO(2,2) \ref\aw{A. Ach\'ucarro and P. K. Townsend, \ajou
Phys. Lett. & 180B (86) 89; \ajou Phys. Lett. & 229B (89) 383; E. Witten,
\ajou Nucl. Phys. & B311 (88) 46; \ajou Nucl. Phys. & B323 (89) 113.}).
Inserting the condition that all fields be independent of $\t$
into the equations of
motion, we obtain two first integrals,
\eqn\bcd{2 {e^a}_\t \omega_{a\t}\qquad{\rm and}
\qquad\Lambda  {e^a}_\t e_{a{\theta}} +
{\omega^a}_\t \omega_{a{\theta}},}
($a = 0,1,2$). The first of these is precisely \twelve, and is equal to
the angular momentum of the spacetime. The second quantity is the first
integral of the $tt$ component of Einstein's equations, and may be identified
with the mass of the spacetime
\ref\fff{The Anti de Sitter group SO(2,2) is a direct
product of two SO(2,1) factors, each of which gives rise to a conserved
charge $M\pm\sqrt{\L}J$. It is interesting to note that
the extremal black holes are solutions for which
one of these charges is zero.}.
A metric like \eleven~has ${e^2}_\t = \Phi$, and
in the gauge ${e^0}_\t={e^1}_\t = 0$,
\eqn\dd{2\Phi \omega_{2\theta} = J, \qquad
\Lambda \Phi^2 + {\omega^a}_\t\omega_{a\theta}= M.}
The BTZ black hole \six~has
\eqna\ee
$$\eqalignno{
&
e^0 = Ndt, \qquad
e^1 = {dr \over N}, \qquad
e^2 = \Phi(N^{\theta}dt + d\theta), & \ee a \cr
&
\omega^0 = \left[ -\half\Phi N {N^\theta}_{,r}-\Phi' N N^\theta\right]
dt-\Phi'Nd\theta,\cr
&
\omega^1 = {-\Phi{N^{\theta}}_{,r} dr\over 2N}, \cr
&
\omega^2=\left[NN_{,r}+\half\Phi^2N^{\theta}{N^{\theta}}_{,r}\right]dt
+\half\Phi^2{N^{\theta}}_{,r}d\theta. & \ee b \cr}
$$
and in that case, given that $\Phi = r$, the form of the functions $N$ and
$N^\t$ follow from equation \dd:
\eqna\ppp
$$\eqalignno{
{N^{\theta}}_{,r}
& = {J\over \Phi^3}={J\over r^3}& \ppp a\cr
- N^2 & = {M -\L\Phi^2 -J^2/4\Phi^2\over (\Phi_{,r})^2} =
M - \L r^2 -J^2/4r^2,& \ppp b
\cr}
$$
up to an irrelevant integration constant.

The effect of the shift
\bb~on $N$ and $N^\theta$
may be computed immediately from
\dd~in a similar way,
by replacing $\Phi$ by $\Phi+ {\lambda/
\Lambda}$, taking the limit $\L\to 0$ and then setting $\Phi=r$
\ref\fn{Note that the angular momentum constraint is replaced by
the condition $\omega_{2\theta}= 0$, irrespective of the value of
$J$. This explains why the action \aa~does not include contributions from
the non-diagonal terms in the metric \eleven, since such contributions appear
multiplied by
$\omega_{2\theta}$, and therefore vanish in the $\Lambda = 0$ limit.}.
It follows that
\eqna\hh
$$\eqalignno{
{N^{\theta}}_{,r}
& = \lim_{\Lambda \rightarrow 0}~ \left[{J \over \left(r + {\lambda
\over \Lambda} \right)^3}\right]=0, & \hh a \cr
- N^2 & = \lim_{\Lambda \rightarrow 0}~\left[M -
\Lambda \left(r + {\lambda \over \Lambda}\right)^2
-{J^2\over 4\left(r+{\lambda\over\L}\right)^4}
\right] =\left(M - {\lambda^2
\over \Lambda}\right) - 2\lambda r\cr
& =  m -2\lambda r. & \hh b \cr}
$$
The dependence on $J$ disappears when the limit is taken, and the solutions
depend only on the parameter $m$.

Following Ref. \ana, the two dimensional metric, determined in this case only
by $N$, is $ds^2 = -N^2 dt^2 + {dr^2 / N^2}$. This is flat,
since $dr /N
= dN / \lambda$. The familiar black hole solution, with a mass
$ m =  M- {\lambda^2 / \Lambda}$, appears when we consider the ``string''
metric, $\Phi^{-1}h_{ij}$:
\eqn\jj{ds^2 = {2\lambda \over
m+N^2} \left[-N^2 dt^2 + {dN^2 \over \lambda^2}\right] }
and can be brought to a more familiar form by the co-ordinate change
$\lambda T = N{\rm  sinh} (\lambda t)$,
$\lambda X = N{\rm  cosh} (\lambda t)$.
Note that the shift relates two dimensional black holes with finite mass $m$ to
three dimensional ones with infinite mass $M$, for which the horizon has been
pushed to infinity and all that remains is the black hole interior, and
vice versa.
\jj~has been extensively studied in \hs, \wbh, \hv~and
we shall not discuss it further, except to say that
the linear dilaton vacuum
solution occurs when $m=0$ or $M=\lambda^2/\L$.

\bigskip
\bigskip
\centerline{\bf Acknowledgements}
We are grateful to M. Crescimanno, D. Cangemi, G. Gibbons, C. Teitelboim
and J. Zanelli for helpful comments.

\listrefs
\vfill\eject

\noindent{\bf Fig. 1:} Regions of the two
dimensional spacetime with $J=0$ covered by the co-ordinates $r,t$,
embedded in Anti-De Sitter space. The diagonal and dotted lines have no
special significance. Here $r_H=\sqrt{M/\L}$.
\medskip
\noindent{\bf Fig. 2:} The Penrose diagram for the region of the
$J=0$ spacetime for which $\Phi\ge 0$. The diagonal lines now represent
event horizons, and the dotted line the ``strong coupling'' region.
\medskip
\noindent{\bf Fig. 3:} The Penrose diagram for the region
of the extremal two dimensional
spacetime with $M=J=0$, for which $\Phi\ge 0$. The dotted line represents
the ``strong coupling'' region.
\medskip
\noindent{\bf Fig. 4:} The Penrose diagram for the two dimensional
spacetimes with $J\ne 0$. (a) has $M\ge\sqrt{\L}|J|$ and
$r_\pm=\left(M\pm\sqrt{M^2-\L J^2}\right)/2\L$, and (b) is the
extremal solution with $M=\sqrt{\L}|J|$ and $r_H=\sqrt{M/2\L}$.

\end